\documentclass[conference]{IEEEtran}
\IEEEoverridecommandlockouts
\usepackage{cite}
\usepackage{tikz}
\usepackage{pgfplots}
\usetikzlibrary{pgfplots.groupplots,patterns}
\pgfplotsset{compat=1.3}
\usepgfplotslibrary{fillbetween}
\usepackage{amsmath,amssymb,amsfonts}

\DeclareMathOperator*{\argmin}{argmin}
\usepackage{algorithmic}
\usepackage[linesnumbered,ruled,vlined]{algorithm2e}
\usepackage{graphicx}
\usepackage{textcomp}
\usepackage{xcolor}
\usepackage{url}
\def\BibTeX{{\rm B\kern-.05em{\sc i\kern-.025em b}\kern-.08em
    T\kern-.1667em\lower.7ex\hbox{E}\kern-.125emX}}

\usepackage{csquotes}
\usepackage{siunitx}
\DeclareSIUnit{\million}{\text{million}}
\DeclareSIUnit{\bit}{bit}
\DeclareSIUnit{\bits}{bits}
\DeclareSIUnit{\siTimes}{\ensuremath{\times}}

\usepackage{subcaption}

\usepackage{xspace}
\xspaceaddexceptions{()}
\usepackage[acronyms]{glossaries}

\newglossaryentry{vdd}{
	name={\ensuremath{V_{\text{DD}}}},
	description={supply voltage},
	symbol=Vdd,
	sort=vdd}
\newglossaryentry{vth}{
	name={\ensuremath{V_{\text{th}}}},
	description={threshold voltage},
	symbol=Vth,
	sort=vth}
\newglossaryentry{gnd}{
	name={GND},
	description={El. ground},
	symbol=GND}

\newglossaryentry{ntype}{
	name={n-type},
	description={pnp doted transistor. Other known names are nmos or nfet}}
\newglossaryentry{ptype}{
	name={p-type},
	description={npn doted transistor. Other known names are pmos or pfet}}
\newglossaryentry{ion}{
	name={\ensuremath{I_{\text{ON}}}},
	description={Current through transistors when it fully conducts},
	symbol=Ion,
	sort=ion}
\newglossaryentry{ioff}{
	name={\ensuremath{I_{\text{OFF}}}},
	description={Current through transistor when it is fully closed},
	symbol=Ioff
	sort=ioff}

\newglossaryentry{vds}{
	name={\ensuremath{V_{\text{DS}}}},
	description={Voltage difference between drain and source contact of a transistor},
	sort=vds}

\newglossaryentry{bitline}{name={bit line},description={Wires that connect columns of \acrshort{sram} cells}}

\newglossaryentry{q}{
	name={Q},
	description={Left inner node of a \acrshort{sram} cell},symbol=Q}
\newglossaryentry{qb}{
	name={QB},
	description={Right inner node of a \acrshort{sram} cell},symbol=QB}

\newacronym{dnn}{DNN}{Deep Neural Network}
\newacronym{tpu}{TPU}{Tensor Processing Unit}
\newacronym{hdc}{HDC}{Hyperdimensional Computing}
\newacronym{hd}{HD}{Hyperdimensional}
\newacronym{am}{AM}{Associative Memory}
\newacronym{im}{IM}{Item Memory}
\newacronym{nvm}{NVM}{non-volatile memory}

\newacronym{ml}{ML}{Match Line}
\newacronym{sl}{SL}{Select Line}
\newacronym{wl}{WL}{Word Line}
\newacronym{bl}{BL}{Bit Line}
\newacronym{blb}{BLB}{Bit Line Bar}
\newacronym{slb}{SLB}{Select Line Bar}

\newacronym{fdsoi}{FD-SOI}{Fully Depleted Silicon-On-Insulator}
\newacronym{box}{BOX}{buried oxide}
\newacronym{sram}{SRAM}{Static Random Access Memory}

\newacronym{fet}{FET}{Field Effect Transistor}
\newacronym{mosfet}{MOSFET}{Metal Oxide Semiconductor \acrlong{fet}}
\newacronym{fefet}{FeFET}{Ferroelectric \gls{fet}}
\newacronym{fefinfet}{Fe-FinFET}{Ferroelectric \acrshort{finfet}}
\newacronym{ic}{IC}{Integrated Circuit}
\newacronym{ram}{RAM}{Random Access Memory}
\newacronym{cam}{CAM}{Content Addressable Memory}
\newacronym{tcam}{TCAM}{Ternary Content Addressable Memory}
\newacronym{cmos}{CMOS}{Complementary MOS}
\newacronym{adc}{ADC}{Analog Digital Converter}
\newacronym{tdc}{TDC}{time-to-digital converter}
\newacronym{mlc}{MLC}{multi-level cell}

\newacronym{fe}{FE}{ferroelectric}

\newacronym{finfet}{FinFET}{Fin \acrlong{fet}}
\newacronym{nc}{NC}{Negative Capacitance}
\newacronym{ncfet}{NCFET}{\acrlong{nc} \acrlong{fet}}

\newacronym{spice}{SPICE}{Simulation with Integrated Circuit Emphasis}

\newacronym{bsim}{BSIM}{Berkeley Short-channel IGFET Model}
\newacronym{cmg}{CMG}{Common Multi Gate}

\glsunset{spice}
\glsunset{mosfet}
\glsunset{cmos}
\glsunset{sram}

\glsunset{hd}
\glsunset{slb}
\glsunset{blb}
\glsunset{fet}
\glsunset{finfet}

\newcommand{\burox}{\gls{box}\xspace}

\newcommand{\ml}{\gls{ml}\xspace}
\newcommand{\spice}{\gls{spice}\xspace}

\newcommand{\sram}{\gls{sram}\xspace}

\newcommand{\ion}{\gls{ion}\xspace}

\newcommand{\vth}{\gls{vth}\xspace}

\newcommand{\iinit}{\ensuremath{I_{\text{init}}}\xspace}

\newcommand{\cmem}{\ensuremath{\text{C}_{\text{mem}}}\xspace}

\newcommand{\xnor}{XNOR\xspace}

\newcommand{\xmax}{\ensuremath{x_\text{max}}\xspace}
\newcommand{\tfire}{\ensuremath{t_\text{fire}}\xspace}
\newcommand{\fmac}{\ensuremath{F_\text{MAC}}\xspace}
\newcommand{\smac}{\ensuremath{S_\text{MAC}}\xspace}

\newcommand{\smacmin}{\ensuremath{S_{\text{MAC,min}}}\xspace}
\newcommand{\sfire}{\ensuremath{S_\text{FIRE}}\xspace}
\newcommand{\sfiremin}{\ensuremath{S_\text{FIRE,min}}\xspace}
\newcommand{\sVfiremin}{\ensuremath{S^{V}_\text{FIRE,min}}\xspace}

\usepackage{cleveref}
\crefname{enumi}{Step}{Steps}
\crefname{section}{Sec.}{Sec.}
\crefname{subsection}{Sec.}{Sec.}
\crefname{figure}{Fig.}{Fig.}
\crefname{algocf}{Algorithm}{Algorithm}
\crefname{algorithm}{Alg.}{Algorithm}
\crefname{algocf}{Algorithm}{Algorithm}
\crefname{table}{Table}{Tables}
\crefname{equation}{Eq.}{Eq.}
\crefname{eqnarray}{Eq.}{Eq.}
\crefname{appendix}{Section}{Sections}

\SetKwInput{KwInput}{Input}                
\SetKwInput{KwOutput}{Output}              
\SetKwRepeat{Do}{do}{while}

\definecolor{1F77B4}{HTML}{1F77B4}
\definecolor{FF7F0E}{HTML}{FF7F0E}
\definecolor{2CA02C}{HTML}{2CA02C}
\definecolor{D62728}{HTML}{D62728}
\definecolor{9467BD}{HTML}{9467BD}
\definecolor{8C564B}{HTML}{8C564B}
\definecolor{E377C2}{HTML}{E377C2}
\definecolor{7F7F7F}{HTML}{7F7F7F}
\definecolor{BCBD22}{HTML}{BCBD22}
\definecolor{17BECF}{HTML}{17BECF}
\definecolor{000000}{HTML}{000000}
\definecolor{FFA280}{HTML}{FFA280}
\definecolor{00C8C8}{HTML}{00C8C8}
\definecolor{B30000}{HTML}{B30000}
\definecolor{800000}{HTML}{800000}
\definecolor{B22222}{HTML}{B22222}
\definecolor{808080}{HTML}{808080}
\definecolor{3399FF}{HTML}{3399FF}
\definecolor{0066FF}{HTML}{0066FF}
\definecolor{0000FF}{HTML}{0000FF}
\definecolor{0000CC}{HTML}{0000CC}
\definecolor{CC99FF}{HTML}{CC99FF}
\definecolor{CC66FF}{HTML}{CC66FF}
\definecolor{CC00FF}{HTML}{CC00FF}
\definecolor{9900CC}{HTML}{9900CC}

\definecolor{66c2a5}{HTML}{66c2a5}
\definecolor{fc8d62}{HTML}{fc8d62}
\definecolor{8da0cb}{HTML}{8da0cb}
\definecolor{377eb8}{HTML}{377eb8}

\begin{document}

\title{
HW/SW Codesign for Robust and Efficient Binarized SNNs by Capacitor Minimization
}

\author{
Mikail Yayla, 
Simon Thomann, 
Ming-Liang Wei, 
Chia-Lin Yang,\\
Jian-Jia Chen,
and Hussam Amrouch 
\IEEEcompsocitemizethanks{
\IEEEcompsocthanksitem Corresponding authors: Mikail Yayla and Hussam Amrouch
\IEEEcompsocthanksitem
M. Yayla and J.-J. Chen are with the Design Automation for Embedded Systems Group, TU Dortmund University, Germany, and Lamarr Institute for Machine Learning and Artificial Intelligence, Germany. Email: \{mikail.yayla, jian-jia.chen,\}@udo.edu
\IEEEcompsocthanksitem
S. Thomann and H. Amrouch are with the Chair of AI Processor Design, Technical University of Munich (TUM) and with the Munich Institute of Robotics and Machine Intelligence (MIRMI). Email: amrouch@tum.de \{thomann, amrouch\}@tum.de.
\IEEEcompsocthanksitem
M.-L. Wei and C.-L. Yang are with the Department of Computer Science and Information Engineering, National Taiwan University (NTU). Email: d04943004@ntu.edu.tw, yangc@csie.ntu.edu.tw.

}}

\maketitle

\begin{abstract}
Using accelerators based on analog computing is an efficient way to process the immensely large workloads in Neural Networks (NNs).
One example of an analog computing scheme for NNs is Integrate-and-Fire (IF) Spiking Neural Networks (SNNs).
However, to achieve high inference accuracy in IF-SNNs, the analog hardware needs to represent current-based multiply-accumulate (MAC) levels as spike times, for which a large membrane capacitor needs to be charged for a certain amount of time.
A large capacitor results in high energy use, considerable area cost, and long latency,
constituting one of the major bottlenecks in analog
IF-SNN implementations.

In this work, we propose a HW/SW Codesign method, called CapMin, for capacitor size minimization in analog computing IF-SNNs.
CapMin minimizes the capacitor size by reducing the number of spike times needed for accurate operation of the HW, based on the absolute frequency of MAC level occurrences in the SW.
To increase the operation of IF-SNNs to current variation, we propose the method CapMin-V, which trades capacitor size for protection based on the reduced capacitor size found in CapMin.
In our experiments, CapMin achieves more than a 14$\times$ reduction in capacitor size over the state of the art, while CapMin-V achieves increased variation tolerance in the IF-SNN operation, requiring only a small increase in capacitor size.


\end{abstract}

\section{Introduction}

The success of neural networks (NNs) brings benefits to numerous fields, while progressively pervading all aspects of our life.
However, the high inference accuracy comes at the cost of large resource demands.
NNs require a massive number of parameters and an immensely high number of multiply-accumulate (MAC) operations.
This poses a profound challenge, because high-performing NN models are becoming increasingly larger, while low-energy operation for sustainability is rapidly gaining importance in a wide range of application domains, especially for embedded systems and edge AI.


{It has been shown that performing the computations of NNs in the analog domain by using Ohm's and Kirchhoff's laws can achieve high resource efficiency~\cite{chi/etal/2016,shafiee/etal/2018}.
}
An example of a computing scheme that exploits this for efficiency is Integrate-and-Fire (IF) Spiking Neural Networks (SNNs)~\cite{wei/etal/2021}.
In IF-SNNs, neural activity is event-driven and described by the integration of current over a certain amount of time, for which a capacitor is used.
If the charge in the capacitor becomes high, a predetermined threshold potential is exceeded, causing the {firing} of an output spike.
The IF-SNN operations use efficient coding and enable efficient operation, because simple analog components can be employed.
Yet, analog computing-based IF-SNNs suffer from nonidealities and variations.
For example, analog multipliers or other components exhibit noisy behavior due to process variation or factors such as temperature.
Furthermore, the capacitor may be too small for correct operation, and analog comparators have limited gain causing undefined output signals for connected digital components.

In particular, the capacitor size is a major bottleneck in the circuit design of IF-SNNs, leading to high energy, area, and latency cost~\cite{xiang/etal/2020, dutta/etal/2020, mingliang/etal/2021}.
Furthermore, it determines the tolerance of the system to nonidealities or variations caused by inherent and external factors.
To the best of our knowledge, principled approaches for minimizing the capacitor size, especially considering nonidealities or variations, do not exist.
The work in~\cite{dutta/etal/2020} compares several implementations of circuits for spike-based inference, where the capacitor size is determined empirically under the engineering constraints.
The studies in~\cite{mingliang/etal/2021, wei/etal/2021} sweep the capacitor size and pick the one satisfying the accuracy constraint.

\begin{figure*}[t!]
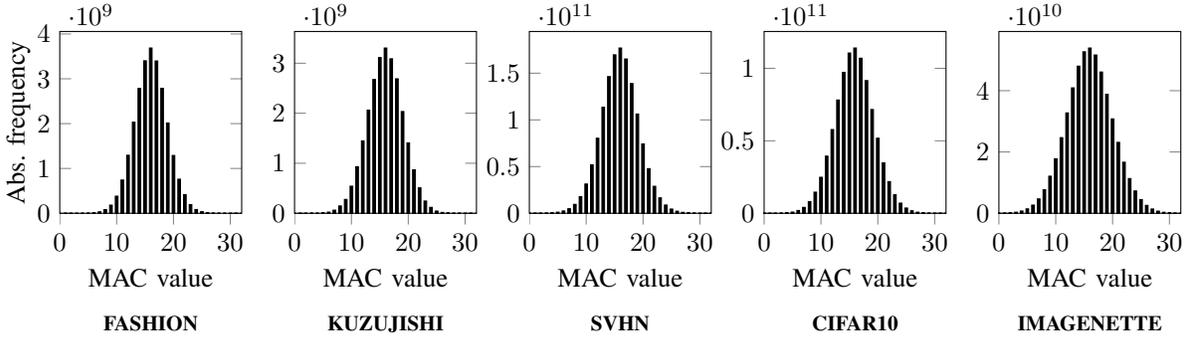

    \begin{center}\scalebox{1}{
        \begin{tikzpicture}
            \begin{groupplot}[group style = {group size = 5 by 1, horizontal sep = 20pt, vertical sep = 42.5 pt}]
            \input{figures/histograms/fmnist}
            \input{figures/histograms/kmnist}
            \input{figures/histograms/svhn}
            \input{figures/histograms/cifar10}
            \input{figures/histograms/imagenette}
            \end{groupplot}
        \end{tikzpicture}}
        \caption{
        {Absolute frequencies of MAC value occurrences (summed over layers) for the training sets. The details of the BNN models are in \cref{tab:expsetting}.}}
        \label{fig:histograms}
    \end{center}
\end{figure*}

These approaches do not use the insights in the SW (NN models), to optimize the HW (IF-SNN circuits). 
In the IF-SNN HW, spike times are required to represent the MAC values that occur with low frequency, wasting valuable capacitor size, in turn leading to high resource cost.
{In this work, we focus on the capacitor size reduction in IF-SNNs executing Binarized Neural Networks (BNNs)~\cite{hubara/etal/2016}.
Due to their binary nature, BNNs are highly resource efficient and robust to variations
\cite{buschjaeger/etal/2021},
making them excellent candidates for analog-based computing with IF-SNNs.
We reveal that many of the MAC values in the BNN SW have a low probability to occur during inference (also in NNs with higher precision regarding weights and inputs, see~\cite{zhao/etal/2019, ding/etal/2019})}.
The lowest and the highest MAC values occur five to seven orders of magnitude less frequently compared to the MAC value at the mean (see histograms in \cref{fig:histograms} for five benchmarks). According to the results shown in \cref{fig:histograms}, the histograms of the MAC values are normally distributed, with a sharp peak at the mean.


\vspace{0.125cm}
\noindent \textbf{Our contributions}: The key focus of our work is exploring HW/SW Codesign methods to achieve IF-SNN operation with a small capacitor, leading to efficient operation through reductions in energy, area, and latency.
We focus on BNNs, which are highly efficient and robust to variations, making them excellent candidates to be executed with IF-SNN HW.
The insights and methods gained by researching
BNNs in this study may also be applicable to higher-precision NNs, if a significant portion of the MAC values in these models have a low probability to occur as well.
Specifically, our contributions are the following:

\begin{itemize}
    \item We propose a novel HW/SW Codesign method called CapMin, which minimizes the capacitor size by reducing the number of spike times needed for accurate operation of the analog IF-SNN HW, based on the absolute frequency of MAC level occurrences in the SW.
    Furthermore, for protection against nonidealities and variations that commonly occur in analog computing, we propose CapMin-V,
    which increases the protecting time margins between the spike times by trading off with capacitor size.
    \item In the experiments, applying CapMin reduces the capacitor size by 14$\times$, leading to significant reductions in resource demand, at a maximum of 1\% accepted accuracy degradation.
    CapMin-V achieves variation tolerant IF-SNN operation by only requiring a $28\%$ increase of the reduced capacitor size from CapMin.
    Our framework for running the experiments is available as open-source in \url{https://github.com/myay/SPICE-Torch}.
\end{itemize}

\newcommand\headingdistHisto{-0.6}
\newcommand\figureWidthHisto{4}
\newcommand\figureHeightHisto{4}

\section{System Model}

We first introduce Binarized Neural Networks in \cref{subsec:bnns} and the operation of IF-SNNs in \cref{subsec:ifsnn}.
Then, we explain the basics and the role of of capacitors in IF-SNNs in \cref{subsec:cap}.
Finally we present the problem definition in \cref{subsec:problem}.

\begin{figure}[t]
    \begin{center}
        \includegraphics[width=1\columnwidth]{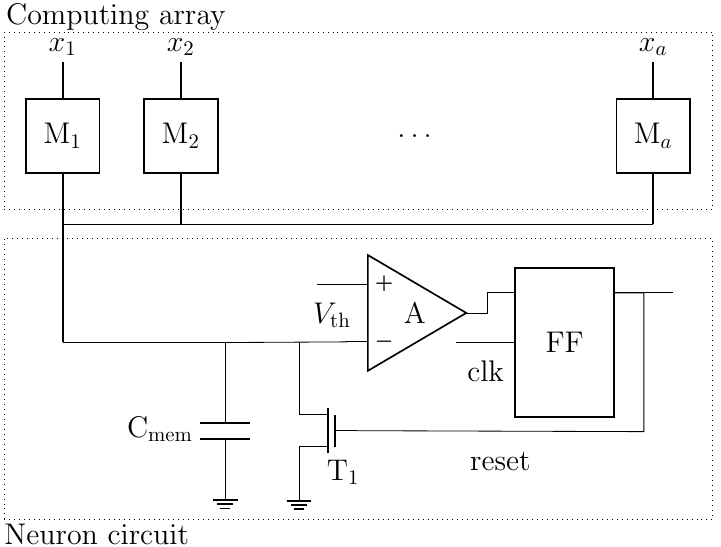}\\%
    \caption{
    {IF-SNN circuit. Top: Computing array with multipliers $M_1$ to $M_a$. Bottom: Neuron circuit.}}
    \label{fig:neuroncircuit}
    \end{center}
\end{figure}

\subsection{Binarized Neural Networks (BNNs)}
\label{subsec:bnns}

In BNNs, the weights and activations are binarized.
The output of a BNN layer can be computed with
\begin{equation}
    2*{{popcount}}({XNOR}(\mathbf{W},\mathbf{X})) - \#{bits} > \mathbf{T},
    \label{eq:bnnformula}
\end{equation}
where ${XNOR}(\mathbf{W},\mathbf{X})$ computes the XNOR operation of the rows in $\mathbf{W}$ with the columns in $\mathbf{X}$ (analogue to matrix multiplication), $popcount$ counts the number of set bits in the XNOR result, $\#bits$ is the number of bits in the XNOR operands, and $\mathbf{T}$ is a vector of threshold parameters that are learned in BNN training, with one entry for each neuron.
The thresholds are computed using the batch normalization parameters, i.e. $T = \mu - \frac{\sigma}{\psi}\eta$, where each neuron has a mean $\mu$ and a standard deviation $\sigma$ over the result of the left side of Eq.~\eqref{eq:bnnformula}, and $\psi$ and $\eta$ are learnable paramaters (details about the batch normalization paramters can be found in~\cite{hubara/etal/2016,sari/etal/2019}).
Finally, the comparisons against the thresholds produce binary values.
In this work, we focus on the computations of the left side of~\cref{eq:bnnformula}, specifically the computation of the popcount result.
In the following, we explain how the popcount can be calculated using the analog based hardware of IF-SNNs.

\subsection{Operation of Binarized IF-SNNs}
\label{subsec:ifsnn}

The circuit of binarized IF-SNNs
using a computing array and the neuron circuit
is shown in \cref{fig:neuroncircuit}, it is based on the work in~\cite{tang/etal/2015}.
In the computing array, $a$ is the array size, $x_{i}$ the input spikes, and $\text{M}_i$ the XNOR gates.
To realize the XNOR, different techniques can be used, e.g.
in the analog domain using
Ohm's law~\cite{mingliang/etal/2021}.
The neuron circuit consists of a membrane capacitor $\text{C}_{\text{mem}}$ with capacitance $C$, an analog comparator A, and a flip flop (FF).
The steps for computations of the MAC results in SNNs are as follows:


(1)
The XNOR gates are loaded with the correct weights or are assumed to be already loaded.
Then the inputs $x_{i}$ are provided to all multipliers in parallel.
The multiplications are all computed in parallel as well,
e.g. employing digital circuits or Ohm's law with memristors~\cite{shafiee/etal/2018}.


(2)
In the neuron circuit, the current charges \cmem.
Once the voltage across \cmem reaches the threshold voltage \vth, an output spike is generated with the analog comparator.
The spike time \tfire is acquired by a counter that tracks the clock cycles until the FF latches the spike signal.


(3)
The spike time is converted to a MAC value by
$\frac{v}{\tfire} = \sum^a_{i=1} w_i x_i,$
where $v=\xmax \frac{C \vth}{\ion}$ and \ion is the on-state current of the multiplier.
The conversion can be described by mappings between sets.
Consider the set of spike times $\sfire = \{t_1, t_2, \dots, t_L\}$, where $t_L$ is the largest firing time, and $t_j < t_{j+1}$.
Consider also the set of MAC-values $\smac = \{q_1, q_2, \dots, q_{L}\}$, where $q_j \leq q_{j+1}$.
In the relation $\sfire \rightarrow \smac$, the values are mapped using $m_j: t_j \rightarrow q_{L-j+1}$.
We organize the index of $q_{L-j+1}$ in a reversed manner to describe the reciprocal relationship between the spike time and the MAC value.
In the state of the art, $L$ is chosen such that each MAC value has a unique spike time.
After completing the calculations, the neuron is reset by the transistor $T_1$.
Because of the limited computing array size, a large vector product (with dimension higher than $a$) is separated into multiple smaller vector products, requiring digital addition for accumulation.
Digital components (adder, reciprocal unit, counter, etc.) are not shown.
They follow conventional designs and
and are not further discussed.

\subsection{Capacitor in IF-SNNs}
\label{subsec:cap}
Capacitors have the capacitance $C=\frac{Q}{V}$ (in \SI{}{\farad} for Farad),
determined by the charge $Q$ placed on the capacitor divided by the voltage $V$ caused by that charge.
A capacitor is charged when a voltage, e.g. $V_0$ is applied,
which causes a current to flow into it.
The charging of a capacitor is described by
\begin{equation}
    V(t)=V_{0} (1-e^{(-\frac{t}{\tau})}),
    \label{eq:capfunc}
\end{equation}
where $V_0$ is the supply voltage, $t$ the time, $\tau=R_{eq}C$ 
the time constant, and $R_{eq}$ the equivalent resistance of the connected circuit from the capacitor's perspective.
No initial charge is in the capacitor.
The capacitor voltage increases rapidly first, but slows down and stops at the maximum capacitor charge $Q=CV_0$.
Since $\tau$ is in the denominator, smaller $C$ lead to larger absolute values in the exponent, in turn causing faster capacitor charging.
In contrast, a larger capacitor leads to slower charging.
The same holds for smaller and larger $R_{eq}$.

\begin{figure}[t]
    \centering
    \begin{tikzpicture}
    \begin{axis}[
      height=5.5 cm,
      width=9 cm,
      ymin=0,
      ymax=0.6,
      xmax=3,
      xmin=0,
      xtick = {0.25, 1.25, 2.25},
      xticklabels={$t_1$, $t_2$, $t_3$},
      yticklabels = {,,},
      ylabel=$V(t)$,
      xtick pos=left,
      ytick pos=left,
      ylabel near ticks,
      ylabel absolute,
      ylabel style={yshift=-1cm},
      legend pos=south east,
      legend style={nodes={scale=0.8, transform shape}, at={(0.74,0.275)},anchor=west},
      ]
      \addplot[color=CC66FF,line width=1.25pt][domain=0:3] {(1-exp(-(x*3))};
      \addplot[line width=1.25pt][domain=0:3] {(1-exp(-x/2))};
      \addplot[color=377eb8,line width=1.5pt][domain=0:3] {(1-exp(-(x/4))};

      \draw [color=2CA02C,line width=1pt] (10,400)-- node[above, xshift=-2cm, yshift=-0.05cm] {\vth} (290,400);

      \def\cheight{525}
      \draw [color=gray] (25,0) -- (25,\cheight) -- (75,\cheight)
      -- (75,0);
      \draw [color=gray] (125,0) -- (125,\cheight) -- (175,\cheight)
      -- (175,0);
      \draw [color=gray] (225,0) -- (225,\cheight) -- (275,\cheight)
      -- (275,0);
      \legend{
      $3\iinit$,
      $0.5\iinit$,
      $0.25\iinit$
    }
    
    \draw[thick](17.5,400) circle (3pt);
    \draw[thick](102,400) circle (3pt);
    \draw[thick](203,400) circle (3pt);
    
    \draw[-, dashed] (17.5,400) -- (17.5, 50);
    \draw[->, dashed] (17.5,50) -- (25,50);
    
    \draw[-, dashed] (102,400) -- (102,50);
    \draw[->, dashed] (102,50) -- (125,50);
    
    \draw[-, dashed] (203,400) -- (203,50);
    \draw[->, dashed] (203,50) -- (225,50);
    
    \end{axis}
\end{tikzpicture}%
    \vspace{-0.5cm}
    \caption{{Voltage across capacitor over time, based on different initial currents. $t_1$, $t_2$, $t_3$ are spike times recorded by clock. Rectangle signal: Clock. Circled points: Ideal spike times.}}
    \label{fig:spiketimes}
\end{figure}
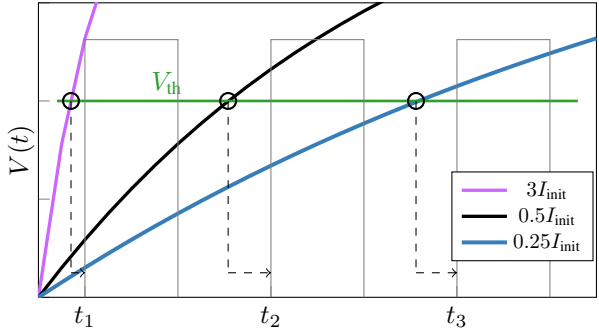

In the IF-SNN circuit, $R_{eq}$ plays an important role.
It depends on the total resistance of all multipliers.
The multipliers can have high or low resistance states (based on the programmed weights).
Due to this, $R_{eq}$ determines the size of the initial current \iinit that flows into the capacitor.
As the current is the first order derivative of the charge with respect to time, i.e., $I(t) = \frac{d Q}{d t}$, by adopting \cref{eq:capfunc} for derivation, we have
$I(t) = C\frac{d V(t)}{d t} = \frac{V_0}{R_{eq}} e^{-\frac{t}{\tau}}$.
Thus, $\iinit = \frac{V_0}{R_{eq}}$ is the largest current.
When the capacitor is fully charged ($t=\infty$), no current flows.
With \iinit and $V_0$, the resistance of the computing array is
$R_{eq} = \frac{V_0}{\iinit}$ by Ohm's law.
When inserting $R_{eq}$ into \cref{eq:capfunc}, we get:
\begin{equation}
    V(t)=V_{0} (1-e^{(-\frac{t}{C} \frac{\iinit}{V_0})}).
    \label{eq:capfuncI}
\end{equation}
Therefore, the larger \iinit, the faster the capacitor is charging.
In \cref{fig:spiketimes}, the voltage curves for different \iinit are shown.

\begin{figure}[t]
    \begin{center}
        \includegraphics[width=1\columnwidth]{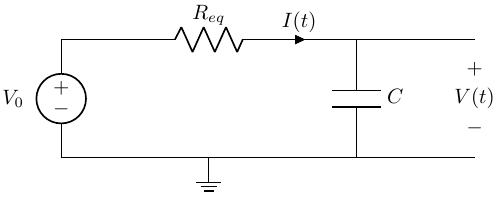}\\%
    \caption{
    {Equivalent representation of IF-SNN circuit in \cref{fig:neuroncircuit}}. $V_0$: Supply voltage. $R_{eq}$: Equivalent resistance of computing array. Voltage $V(t)$ across capacitor is measured over time.}
    \label{fig:eqneuroncircuit}
    \end{center}
\end{figure}

The equivalent RC circuit with $V_0$, $R_{eq}$, $I(t)$, and $V(t)$ is shown in \cref{fig:eqneuroncircuit}.
In the IF-SNN circuit, $R_{eq}$ is the resistance of the computing array, seen from the perspective of the capacitor.
The equivalent resistance $R_{eq}$ of all multipliers determines the size of $I_0$ (and thus $I(t)$)  flowing into the capacitor, causing $V(t)$ to rise.


The charging properties of capacitors are used to realize IF-SNN circuits (\cref{subsec:ifsnn}).
A spike occurs ideally when the charge (integrated current from the computing array over time) in the capacitor leads to $V(t) = \vth$.
The ideal firing times, where the voltage curve and the \vth-line cross, are marked with circles in \cref{fig:spiketimes}.
A spike can only be registered by the FF at the rising edges of the clock (see the gray clock signals in \cref{fig:spiketimes}), therefore the spike only occurs at quantized time points at the rising edges.

With a fixed clock frequency, the size of the capacitor is chosen such that all required firing times in the set \sfire are represented uniquely.
The higher the number of firing times to include, the larger the required capacitor size.
This is further exacerbated by the exponential flattening of the capacitor voltage over time.
Due to this, the distances between subsequent $t_i$ have to increase as well.
Therefore, the spike times in the set \sfire do not include all time points of rising clock edges, but time points at which a spike occurs when known initial currents are applied.

\subsection{Problem Definition}
\label{subsec:problem}

We are given a BNN model and an IF-SNN circuit (see \cref{subsec:ifsnn}) to perform its computations.
The IF-SNN circuit has a capacitor with capacitance $C$, whose behavior is described in \cref{eq:capfunc}.
For every different summed current that flows from the computing array to the capacitor, a unique spike time is placed in \sfire, representing a MAC value in \smac.
The higher the number of spike times used to represent the MAC values, the larger the required capacitor size.
Note that the capacitor size is a major bottleneck in the circuit design of IF-SNNs, as it leads to high cost in energy, area, and latency~\cite{xiang/etal/2020, dutta/etal/2020, mingliang/etal/2021, moitra/etal/2022}.

{In this work, our first goal is to} construct the sets \sfire and \smac, such that the capacitor size is minimized and therefore \emph{energy, area, and latency} of the IF-SNN circuit, while limiting the inference accuracy drop of the BNN.
Our second goal is to modify the sets \sfire and \smac from CapMin, such that the tolerance to process variation, which can significantly affect the correctness of analog computing schemes, is increased.

\section{Our Proposed Methods: CapMin and CapMin-V}
In \cref{subsec:m1}, we propose our method CapMin (Capacitor size minimization), in which spike times in \sfire are only assigned to the most important MAC values.
CapMin does not protect against process variation, therefore, in \cref{subsec:m2}, we present CapMin-V, which aims to achieve variation tolerant IF-SNN operation by trading off with capacitor size.

\label{sec:methods}
\subsection{Our Method CapMin for Capacitor Minimization}
\label{subsec:m1}


We consider that the most important MAC values are the ones that occur most frequently during the inference of NNs.
In \cref{fig:histograms}, we present the histogram of all MAC value occurrences in forward passes with the training set.
From this intuition, we propose a capacitor minimization procedure.
We reduce the number of required spike times in \sfire based on the absolute frequency of the observed MAC value occurrences in \smac.
This reduces the capacitor size.

We denote the MAC values occurring during IF-SNNs operation with $\smac =  \{ q_1, q_2, \dots, q_L \}$.
To these values, spike times in $\sfire = \{ t_1, t_2, \dots, t_L \}$ are assigned bijectively.
There are $L$ mappings $m_j: t_j \rightarrow q_{L-j+1}$ (see \cref{subsec:ifsnn}).
$L$ is initialized such that each MAC value has a unique spike time.
For each MAC value in $\smac$, we extract its absolute frequency of occurrences (AFO), i.e. $\fmac = \{ f_1, f_2, \dots, f_L\}$, where $f_i$ counts the number of occurrences of the MAC value $q_i$.
This is achieved by tracking the MAC values in the computing array (\cref{fig:neuroncircuit}) during the inference of the NN.

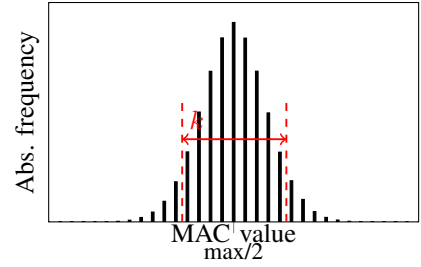
\begin{figure}[t]
    \centering
    \begin{tikzpicture}
    \begin{axis}
        [
            height=4.5 cm,
            width=6.5 cm,
            ymin = 1,
            ybar,
            xmin=0,
            xmax=32,
            xtick = {-1, 16, 33},
            xticklabels={,\small{$\text{max}$/2},},
            title style={at={(0.5,0.975)},anchor=north,yshift=-2},
    		xlabel = MAC value,
    		ylabel = Abs. frequency,
    		ylabel near ticks,
            xtick pos=bottom,
            xlabel style={at={(0.5,0)},yshift=0.1cm},
            ytick style={draw=none},            ytick=\empty
        ]
        \addplot[fill,bar width=1pt,color=000000] table [x, y] {fmnist_hist_accumulated.dat};
    \end{axis}
    
    
    \draw [dashed, color=red, line width = 0.025cm] (1.775,0)--(1.775,1.6);
    \draw [dashed, color=red, line width = 0.025cm] (3.16,0)--(3.16,1.6);
    
    \draw [<->, color=red, line width = 0.025cm] (1.775,1.1) --  node[above,xshift=-0.5cm] {$k$} (3.16,1.1);
\end{tikzpicture}%
    \vspace{-0.25cm}
    \caption{
    {Role of inclusion parameter $k$ in histogram of MAC values. All MAC values within borders get a unique spike time value assigned.
    The larger $k$, the more values within the borders.}}
    \label{fig:histo_method}
\end{figure}

We use the set \fmac to minimize the number of MAC values in \smac, to construct a set \smacmin.
To this end, only the $k$ MAC levels with the highest AFO are added to \smacmin.
MAC values that have low AFO are not added to \smacmin and get mapped to the nearest MAC level in \smacmin.
The role of $k$ and its clipping behavior in the histogram of \fmac is shown in \cref{fig:histo_method}.
The value of $k$ can be configured based on the desired number of MAC levels in \smacmin.
Since the mapping from \smacmin to \sfiremin is bijective, the number of spike times in \sfiremin that are needed to represent the MAC values are limited by $k$.
In turn, less spike times require a smaller capacitor in the neuron circuit, leading to its minimization.

As a result of CapMin, the original set \smac is clipped to the set \smacmin based in the information in \fmac and on $k$.
For clipping the set \smac to \smacmin the following function is used, where the smallest MAC value in \smacmin is $q_{\text{first}}$, $q_{\text{last}}$ the largest, and the MAC value is $M=\sum^a_{i=1} w_i x_i$:

\begin{equation}
    \label{eq:clipping}
        M = \left.
        \begin{cases}
            M, & \text{for } q_{\text{first}} \leq M\leq q_{\text{last}}\\
        q_{\text{first}}, & \text{for } M \leq q_{\text{first}}\\
        q_{\text{last}}, & \text{for } M \geq q_{\text{last}}
        \end{cases}
        \right\}
\end{equation}















\begin{algorithm}[t]
\DontPrintSemicolon

    \KwInput{$\phi$, $\sfiremin = \{t_1, t_2, \dots, t_{k} \}$}
    \KwOutput{\sVfiremin}

    $\sVfiremin \leftarrow \sfiremin$

    $\phi_{\text{step}} \leftarrow 1$, $k_{V} \leftarrow k$

    \While{$\phi_{\text{step}}$ $\leq$ $\phi$}{

        $j \leftarrow \argmin(\operatorname{diag} (P_{\text{map}}))$

        Handle out-of-bound cases

        \If {$p_{j-1,j-1} < p_{j+1,j+1}$}{

            \For {$i$ in $\{1, \dots, k_{V}\}$}{
                $p_{i,j-1} \leftarrow p_{i,j-1} + p_{i,{j}}$
            }
        }
        \Else{
            \For {$i$ in $\{1, \dots, k_{V}\}$}{
                $p_{i,j+1} \leftarrow p_{i,j+1} + p_{i,{j}}$
            }
        }

        Remove column and row $j$ from $P_{\text{map}}$

        Remove $t_j$ from \sVfiremin

        $\phi_{\text{step}} \leftarrow \phi_{\text{step}}+1$, $k_V \leftarrow k_V - 1$
    }

    Add padding to $P_{\text{map}}$

    return \sVfiremin

\caption{CapMin-V: Constructing the set \sVfiremin}
\label{alg:svmacmin}
\end{algorithm}

\subsection{Our Method CapMin-V for Protection against Variations}
\label{subsec:m2}

\begin{figure}[t]
    \centering
    \begin{tikzpicture}
    \begin{axis}[
      height=6 cm,
      width=10 cm,
      ymin=0,
      ymax=0.4,
      xmax=3,
      xtick = {0.25, 1.25, 2.5},
      xticklabels={$t_1$, $t_2$, $t_3$},
      yticklabels = {,,},
      ylabel=$V(t)$,
      xtick pos=left,
      ytick pos=left,
      ylabel near ticks,
      ylabel absolute,
      ylabel style={yshift=-1cm},
      legend pos=north east,
      legend style={at={(1,1)}, nodes={scale=0.7, transform shape}},
      ]
      \addplot[color=brown, line width=0.75pt, name path=f1][domain=0:5] {(1-exp(-x*4))};
      \addplot[color=black,line width=0.75pt][domain=0:5] {(1-exp(-(x/2))};
      \addplot[color=red, line width=0.75pt, name path=f2][domain=0:5] {(1-exp(-x/4.5))};
      
      \addplot[mark=star, color=brown, line width=0.5pt, name path=f3][domain=0:5] {(1-exp(-x/2.5))};
      \addplot[mark=star,color=black,line width=0.5pt][domain=0:5] {(1-exp(-(x/5.25))};
      \addplot[mark=star, color=red, line width=0.5pt, name path=f4][domain=0:5] {(1-exp(-x/10))};

    
    \addplot [
        thick,
        color=red,
        fill=CC00FF, 
        fill opacity=0.15,
    ]
    fill between[
        of=f1 and f2,
        split,
        every segment no 0/.style={
            fill=none,
        },
    ];
    
    \addplot [
        thick,
        color=blue,
        fill=377eb8, 
        fill opacity=0.15,
    ]
    fill between[
        of=f3 and f4,
        split,
        every segment no 0/.style={
            fill=none,
        },
    ];
    
    

    \addplot [
        pattern=north west lines,
        pattern color=black,
    ]
    fill between[
        of=f2 and f3,
        split,
        every segment no 0/.style={
            fill=none,
        },
    ];
    
    
    
      \draw [color=2CA02C,line width=1pt] (-20,20) -- node[above, xshift=-2.75cm, yshift=-0.05cm] {$V_{th}$} (295,20);

      \legend{
      {$I_i + \epsilon_i$},
      {$I_i$},
      {$I_i - \epsilon_i$},
      {$I_{i+1} + \epsilon_{i+1}$},
      {$I_{i+1}$},
      {$I_{i+1} - \epsilon_{i+1}$}
    }
    

    
    

    \def\cheight{35}
    \def\clkW{12.5}

    \foreach \iValue in {25, 50,...,450}{
        \edef\temp{\noexpand\draw [color=gray] (\iValue,0) -- (\iValue,\cheight) -- (\iValue+\clkW,\cheight) -- (\iValue+\clkW,0);}
        \temp
    }
    
    \draw[-, dashed, line width=0.5pt] (75,40) -- node[above, xshift=0cm, yshift=-2.25cm] {$t^{\text{RI}}_1~t^{\text{LI}}_2$} (75,0);
    \draw[-, dashed, line width=0.5pt] (187.5,40) -- node[above, xshift=0cm, yshift=-2.25cm] {$t^{\text{RI}}_2~t^{\text{LI}}_3$} (187.5,0);
    
    \end{axis}
    
\end{tikzpicture}%
    \vspace{-0.25cm}
    \caption{
    {Effect of current variation on capacitor charging. Charging is shown in black for $I_i$ and $I_{i+1}$. Depending on sign of variation ($\epsilon_i$ or $\epsilon_{i+1}$), capacitor may charge faster (brown) or slower (red). Variations can cause any deviation in the purple (for $I_i$) or the blue area (for $I_{i+1}$). Charging curves under variation may overlap (striped area).}}
    \label{fig:spiketimeswithV}
\end{figure}
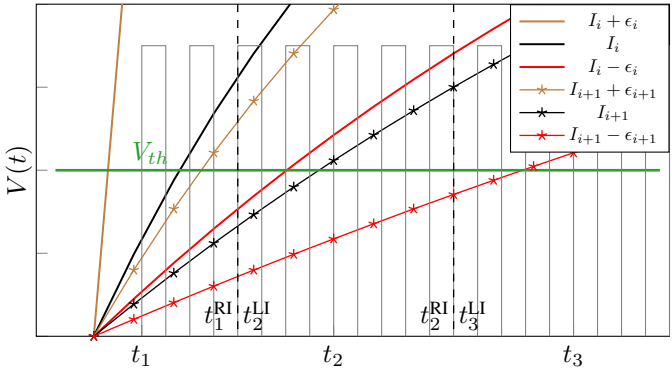

CapMin does not account for process variation.
In fact, process variation can cause current variation, which may affect the correctness of IF-SNNs operations.
Therefore, in the following we analyze the effects of current variation on the IF-SNN operation and based on the analysis propose the method Capmin-V for protection.

We solve \cref{eq:capfuncI} for $t$ when $V(t)=\vth$.
Denoting $I_i$ with index $i$ for the $i$th initial current instead of \iinit, we get
\begin{equation}
    \label{eq:tvi}
    t(I_i) = - \frac{CV_0}{I_i} \ln(1-\frac{\vth}{V_0}),
\end{equation}
where $I_i$ is decreasing with increasing $i$, i.e. $I_i > I_{i+1}$.
In IF-SNNs, $I_i$ leads to the spike time $t_i$ in \sfire.
For example, $I_1$ is the largest current leading to the shortest spike time $t_1$ (mapped to the highest MAC value $q_L$).
$I_L$ is the smallest current leading to the longest spike time (mapped the the smallest MAC value $q_1$).
Since the currents coming out of the XNOR cell are all the same (assuming the same states), the difference between $I_i$ and $I_{i+1}$ is constant: $I_i - I_{i+1} = c>0~\forall i$.

Without variations, the function $t(I_i)$ is deterministic.
It produces the same spike time $t_i$ (in the set \sfire) for a certain $I_i$ (see black plots in \cref{fig:spiketimeswithV}).
If $I_i$ has variations, $t(I_i)$ will also change.
The variations in $I_i$ are proportional to $I_i$ (a certain percentage of it), with a certain mean and variance.
We define the measured maximum of $I_i$ variation as $\epsilon_i$.
Due to $\epsilon_i$ of $I_i$, $t$ may fall into the interval $\mathcal{E}_i = [t(I_i + \epsilon_i), t(I_i - \epsilon_i)]$, where $|\mathcal{E}_i|$ is its length.
In this case, a $t$ that is not in the set \sfire will be calculated in \cref{eq:tvi}.
The result of \cref{eq:tvi} under variations is assigned to the nearest $t_i$ in \sfire, where the midpoints between two spike times are the assignment thresholds.
The assignment threshold on the right of $t_i$ is $t^{\text{RI}}_i = t_i + \frac{t_{i+1} - t_i}{2}$ and $t^{\text{LI}}_i = t_i - \frac{t_{i} - t_{i-1}}{2}$ on the left.
We define the interval $B_i = [t^{\text{LI}}_i, t^{\text{RI}}_i]$ and its length as $|B_i|$.
The interval boundaries are shown in the dashed vertical lines in \cref{fig:spiketimeswithV}.
Any spike time that occurs in $B_i$ is assigned to $t_i$.
If the variation of $I_i$ is large enough to make \cref{eq:tvi} cross the interval borders $t^{\text{LI}}_i$ or $t^{\text{RI}}_i$, $I_i$ will erroneously be assigned to a wrong spike time, e.g. $t_{i-1}$, $t_{i+1}$, or other spike times farther away.
This is shown in the striped area in \cref{fig:spiketimeswithV}.


The probabilities for $t_i$ to assume $t_j$, which may be different than $t_i$ due to current variations, are modeled in the matrix in \cref{eq:probm}.
The first index in $P_{\text{map}}$ describes the spike time that has variations.
The second index describes the erroneous spike time selected due to variations.
For example, $t_1$ has the probability $p_{1,1}$ to assume $t_1$ and $p_{1,2}$ for $t_2$.
If all the diagonal elements are ``1'', then it is equivalent to the direct mapping mapping in \cref{subsec:m1} (ideal case, no variations).
\begin{equation}
    \label{eq:probm}
    P_{\text{map}} =
    \begin{bmatrix}
    p_{1,1} & p_{1,2} &\dots & p_{1,L}\\
    p_{2,1} & p_{2,2} &\dots & p_{2,L}\\
    \dots & \dots &\dots & \dots\\
    p_{L,1} & p_{L,2} &\dots & p_{L,L}
\end{bmatrix}
\end{equation}


To increase variation tolerance, it needs to be known which spike times have lower or higher variation tolerance.
Consider the differences in length between the intervals $\mathcal{E}_i$ and $B_i$.
Due to the increase of $t(I_i)$ with smaller $I_i$, $|B_i|$ gets larger.
In the same way, $|\mathcal{E}_i|$ gets larger as well.
However, the variation $\epsilon_i$ becomes smaller with smaller currents, since $I_{i} - I_{i+1} =c > 0~\forall i$ is assumed to be constant, while
$\epsilon_i$ is proportional to the size of $I_i$.
To conclude, the intervals $B_i$ and $\mathcal{E}_i$ both get larger with increasing $i$, but the variations $\epsilon_i$ become smaller, i.e. as $I_i$ becomes smaller, the ratio $r_i = \frac{|B_i|}{|\mathcal{E}_i|}$ gets larger, with a larger margin for tolerating variations.
Based on this, we hypothesize that the spike times with larger $t_i$ are more tolerant to variations than spike times with smaller $t_i$.

Based on this hypothesis and the error matrix in \cref{eq:probm}, we propose the method CapMin-V to increase the tolerance of the IF-SNNs to variations at the cost of capacitor size.
As the starting point, we use the set \sfiremin with $k$ elements and extract its $P_{\text{map}}$.
We aim to increase the probabilities in $P_{\text{map}}$ for a spike time to assume correct values under current variations.
Therefore, the criterion for optimization is to maximize the individual probabilities on the diagonal, i.e. the $p_{i,i}$.
In CapMin-V, the $p_{i,i}$ are increased by merging the spike times $t_i$ in \sfiremin of the smallest $p_{i,i}$ with the neighboring spike times.
By this way, a spike time with higher $p_{i,i}$ is created, which in turn leads to higher tolerance to current variations since the $r_i$ is increased.

Before merging, $t_i$ and $t_{i+1}$ in \sfiremin are different spike times.
With application of CapMin-V, the time intervals of $t_i$ and $t_{i+1}$ are merged to create a new, variation tolerant spike time.
When the neighboring $t_{i+1}$ (or $t_{i-1}$) is merged with $t_i$, the spike time $t^{+}_{i}$ is created, which has a larger time margin to the subsequent spike time compared to $t_i$.
The new spike time intervals of $t^{+}_{i}$ are $t^{\text{RI+}}_{i+1} = t_i + \frac{t_{i+2} + t_{i+1}}{2}$ and $t^{\text{LI+}}_i = t_i - \frac{t_{i} + t_{i-1}}{2}$ on the left (stays the same).
The interval border to the right of $t^{+}_i$ is larger than the one from $t_i$, since $t_{i+1} + t_{i} < t_{i+2} + t_{i+1}$.
The same holds for merging time $t_{i}$ with $t_{i-1}$, just the different way around.
To merge, the probabilities of two neighboring columns in \cref{eq:probm} need to be added, i.e. $p_{i,j+1} \leftarrow p_{i,j+1} + p_{i,j} \forall j$ for merging $t_{i}$ with $t_{i+1}$, and $p_{i,j-1} \leftarrow p_{i,j-1} + p_{i,j} \forall j$ for merging $t_{i}$ with $t_{i-1}$.
Due to the summing of probabilities, the probabilities on the diagonal are increased.

The procedure of CapMin-V is in \cref{alg:svmacmin}.
\sfiremin (from CapMin) and $\phi$ (nr. of mergings to be performed) are the inputs.
First, \sVfiremin is initialized as \sfiremin.
To maximize the $p_{i,i}$, the minimum $p_{i,i}$ in $P_{\text{map}}$ is determined and the index is stored in $j$.
If $j$ is the right bound, a left merge will be performed, and the other way around for the left bound.
Then the column of the smallest $p_{i,i}$ is merged with a neighboring column.
Whether to merge it left or right is decided by the $p_{i,i}$ of the left or right neighbor.
If $p_{i-1,i-1}$ (diagonal entry of left neighbor) is smaller than $p_{i+1,i+1}$ (diagonal entry of right neighbor), a left merge will be performed.
Otherwise, a right merge will be performed.
Boundary cases are merged to inner directions and ties are broken arbitrarily.
After adding the probabilities,
in $P_{\text{map}}$, the column of the merged spike time $p_{i,i}$ is removed (since it has been added to the neighboring spike time) and its row as well, since the spike time does not occur any more.
Then, $k_V$ is decremented.
After the specified number of mergings $\phi$, the algorithm pads $P_{\text{map}}$ with zeros on the left and right, and adds 1s to realize the clipping from CapMin.
Finally, \sVfiremin is returned and its spike times are mapped to the $k$ most frequently occurring MAC values.

\section{Experiment Setups and Results}

We present the experiment setups in \cref{subsec:setup}.
Then, we evaluate our proposed methods CapMin and CapMin-V in \cref{subsec:capminexp} and \cref{subsec:capminVexp} respectively.

\subsection{Experiment Setups}
\label{subsec:setup}

\begin{table}[t]
  \centering
  \begin{tabular}{l|l|l|l|l}
  Name         & \# Train & \# Test & \# Dim & \# classes       \\ \hline
   FashionMNIST (VGG3) & 60000 & 10000 & (1,28,28) & 10 \\
  KuzushijiMNIST (VGG3) & 60000    & 10000   & (1,28,28)  & 10   \\
  SVHN (VGG7)         & 73257    & 26032   & (3,32,32)  & 10  \\
  CIFAR10 (VGG7)     & 50000    & 10000   & (3,32,32)   & 10  \\
  Imagenette (ResNet18)     & 9470    & 3925   & (3,64,64)   & 10  \\
  \end{tabular}
  \caption{
  {Datasets used for experiments.}}
  \label{tab:datasets}
\end{table}

\begin{table}[t]
  \centering
  \begin{tabular}{@{}l|l@{}}
  Name                                   & Architecture  \\ \hline
  VGG3 & In $\to$ C64 $\to$ MP2  $\to$ C64 $\to$ MP2 $\to$ FC2048  $\to$ FC10 \\[1mm]
  VGG7 & In $\to$ C128  $\to$ C128 $\to$ MP2 $\to$ C256  $\to$ C256 $\to$ MP2  \\
  & \phantom{In} $\to$ C512  $\to$ C512 $\to$ MP2 $\to$ FC1024  $\to$ FC10 \\
  ResNet18 & In $\to$ C64 $\to$ SCB64 $\to$ SCB128 $\to$ SCB256 $\to$ MP2 \\
  & \phantom{In} $\to$ SCB512  $\to$ MP4 $\to$ FC10 \\
  \end{tabular}
  \caption{
  {
  BNNs with fully connected (FC), convolutional (C), and maxpool (MP) layers. SCB: Skip-connection block. Convolutional layers are followed by batch normalization layers (except output).}}
  \label{tab:expsetting}
\end{table}

\begin{figure}[ht!]
    \centering
    \includegraphics[width=1\columnwidth]{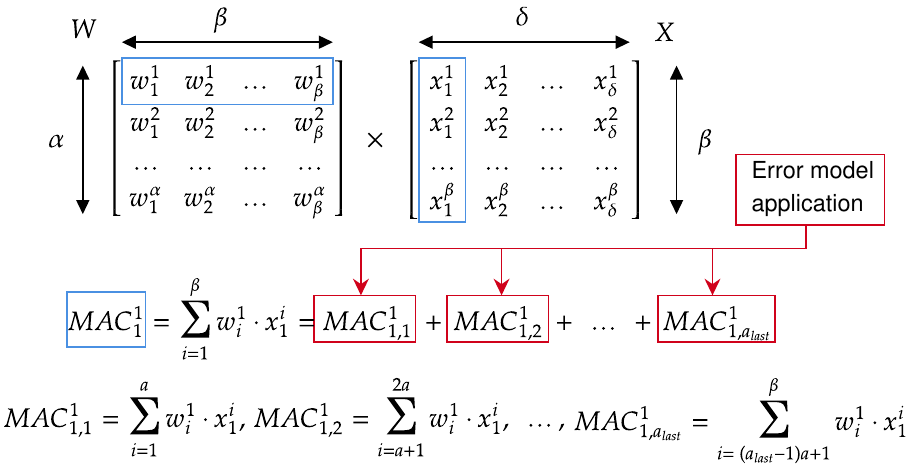}\\%
    \caption{
    {Workload of NNs in matrix notation and error application at the MAC-level. Every convolution in BNNs can be expressed as this matrix product. $\alpha$: Number of neurons, $\beta$: Number of weights, $\delta$: Columns in unrolled input matrix, $a$: Array size.}}
    \label{fig:errorappl}
    \vspace{-0.25cm}
\end{figure}

\newcommand\headingdist{1.05}
\newcommand\figureWidth{7.5}
\newcommand\figureHeight{4.5}

\newcommand\figureWidthA{6}
\newcommand\figureHeightA{4.5}

\begin{figure*}[t!]
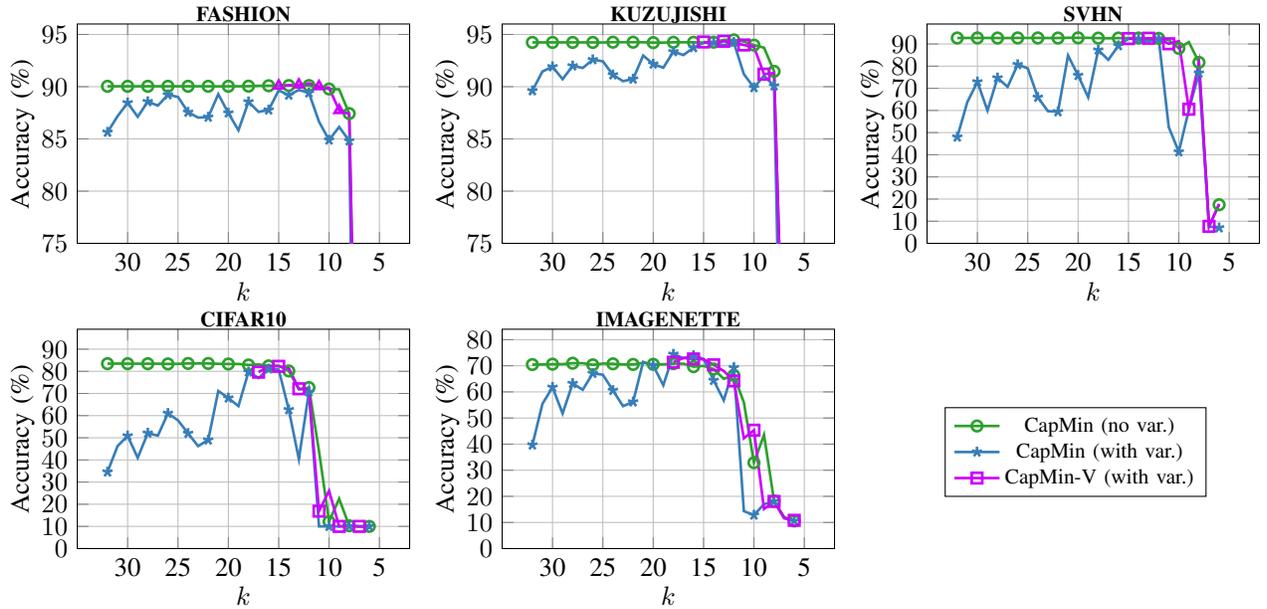

    \begin{center}\scalebox{1}{
        \begin{tikzpicture}
            \begin{groupplot}[group style = {group size = 3 by 2, horizontal sep = 35pt, vertical sep = 32.5pt}]
            \input{figures/accuracy/fmnist}
            \input{figures/accuracy/kmnist}
            \input{figures/accuracy/svhn}
            \input{figures/accuracy/cifar10}
            \input{figures/accuracy/imagenette}
            \end{groupplot}
        \end{tikzpicture}}
        \vspace{-0.1cm}
        \caption{
        {Accuracy over $k$. The higher $k$, the larger capacitor size. Capacitor size range: From \SI{135.2}{\pico\farad} ($k=32$) to \SI{1}{\pico\farad} ($k=5$).}}
        \label{fig:accuracy_results}
        \vspace{-0.25cm}
    \end{center}
\end{figure*}

\newcommand\figureWidthB{4}
\newcommand\figureHeightB{3.75}

\hspace{-1cm}

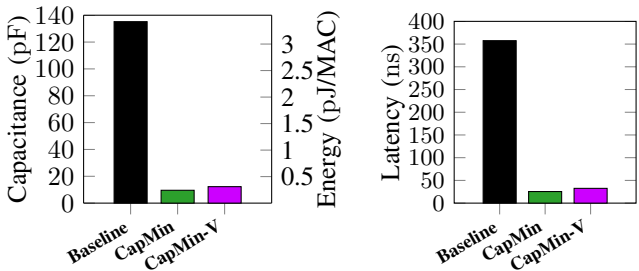
\begin{figure}[t!]
    \begin{center}
        \vspace{-0.5cm}
        \scalebox{1}{
    \begin{tikzpicture}
        \begin{axis}[
            height=2.5 cm,
            width=2.5 cm,
            xmin = 0,
            xmax = 4,
            ymin=0,
            ymax=140,
            ytick distance=20,
            scale only axis,
            ybar,
            bar shift=0,
            bar width=12.5,
            axis y line*=left, 
            ylabel=Capacitance ($\SI{}{{\pico \farad}}$),
            xtick={1,2,3},               
            xticklabels={
            \textbf{Baseline},
            \textbf{CapMin},
            \textbf{CapMin-V}
            },
            xtick pos=bottom,
            x tick label style={rotate=30,anchor=east},xticklabel style={align=right, text width=2cm, font=\scriptsize, yshift={-0.1cm}},
            ylabel shift = -0.2cm,
        ]
        \addplot[draw=black,fill=000000]coordinates{(1,135.2)};
        \addplot[draw=black,fill=2CA02C]coordinates{(2,9.575)};
        \addplot[draw=black,fill=CC00FF]coordinates{(3,12.27)};
        \end{axis}
        \begin{axis}[   
            height=2.5 cm,
            width=2.5 cm,
            scale only axis,
            ymin=0,
            ymax=140,
            ybar,
            axis y line*=right,
            axis x line=none,
            ylabel=Energy ($\SI{}{{\pico \joule}}$/MAC),
            ytick={19.753, 39.506, 59.259, 79.0123, 98.765, 118.519},               
            yticklabels={0.5, 1, 1.5, 2, 2.5, 3},
            ylabel shift = -0.2cm,
            ]
            \addplot [white,draw opacity=0] {x+6};
        \end{axis}
    \end{tikzpicture}%
    \begin{tikzpicture}%
        \begin{axis}[
        	bar width=12.5,
            ybar,
            bar shift=0,
            ymin=0,
            ymax=400,
            xmin = 0,
            xmax = 4,
            ytick distance=50,
            xtick={1,2,3},
            xticklabels={
                \textbf{Baseline},
                \textbf{CapMin},
                \textbf{CapMin-V}
            },
            xticklabel style={text width=2cm,align=center},
            ylabel=Latency ($\SI{}{{\nano \second}}$),
            xlabel near ticks,
            ylabel near ticks,
            ylabel shift = -0.2cm,
            height=4 cm,
            width=4 cm,
            nodes near coords align={vertical},
            x tick label style={rotate=30,anchor=east},
            xticklabel style={align=right, text width=2cm, font=\scriptsize, yshift={-0.1cm}},
            xtick pos=bottom,
            yticklabel style = {font=\small}
        ]
        \addplot[draw=black,fill=000000]coordinates{(1,357.5)};
        \addplot[draw=black,fill=2CA02C]coordinates{(2,25.5)};
        \addplot[draw=black,fill=CC00FF]coordinates{(3,32.5)};
        \end{axis}
    \end{tikzpicture}%
}%
        \vspace{-1.25cm}
        \caption{
        {Capacitor size and latency comparison of the neuron circuit
        for the baseline and our two proposed methods at 1\% accuracy cost.}}
        \label{fig:cap_lat}
        \vspace{-0.4cm}
        \label{fig:test}
    \end{center}
\end{figure}

\subsubsection{BNN Training Setup in PyTorch}
To demonstrate the effectiveness of our proposed methods, we employ BNNs which are executed as IF-SNNs using the hardware configuration in \cref{fig:neuroncircuit}.
We use the datasets Fashion, Kuzujishi, SVHN, CIFAR10, and Imagenette (a subset of ImageNet scaled to $64 \times 64$), see~\cref{tab:datasets}.
We use the training sets to extract \fmac for the methods in \cref{sec:methods} and evaluate the accuracy using the test sets.
Our models (\cref{tab:expsetting}) are modified and binarized (weights and inputs) based on the architectures of VGG and ResNet~\cite{simonyan/etal/2014,he/etal/2016}, adapted for the above datasets.
We use moderately difficult prediction tasks, with a small VGG3 model and relatively large VGG7 and ResNet18 models.
The BNNs are up-to-date, suitably sized (not overparametrized) and capable models tailored for resource-constrained inference.
Note that we use the weakest variant of BNNs, with binarized weights and binarized activations, which are the hardest to train.
The details of the datasets and BNNs are in \cref{tab:datasets} and \cref{tab:expsetting}.
The BNNs use convolutional (C) layers with size $3 \times 3 $, fully connected (FC), maxpool (MP) with size $2 \times 2$, and batch normalization (BN).
We use Adam for optimizing BNNs and the modified hinge loss (MHL) with the hyperparameter $b=128$~\cite{buschjaeger/etal/2021}
to achieve high accuracy and error tolerance by margin-maximization.
The batch size is $256$ ($128$ for Imagenette) and the initial learning rate (LR) is $10^{-3}$ in all cases.
We halve the LR every 10th epoch for Fashion, Kuzujishi, SVHN, and halve it every 50th epoch for CIFAR10 and Imagenette.
For each model we train 100 epochs for Fashion, Kuzujishi, SVHN, and 200 epochs for CIFAR10 and Imagenette.
Note that we do not apply any retraining in this work.
All method are applied on in the post-training stage without any changes to the BNN models.

\subsubsection{SPICE Setup}
In the computing array (\cref{fig:neuroncircuit}), we use \sram-based XNOR cells with \SI{14}{\nano\meter}
FD-SOI
technology,
of which we reproduce industry measurements
with a ultra-thin body and \burox design~\cite{liu2013fdsoi_measurements}.
The transistor model-card parameters for the industry-standard compact model of FD-SOI (BSIM-IMG) are carefully tuned until they are in excellent agreement with the measurements.
The model is also calibrated to device-to-device variation measurements.
For a comprehensive variability representation, all important sources of process variation (gate work function, channel dimension, \burox and channel thickness) are considered.
Through \spice Monte-Carlo simulations based on the calibrated compact model, the standard deviation for each model parameter is tuned to match the observed variation in the measurements.
We use an array of $a=32$ XNOR cells to realize the computing array.
Each XNOR cell connects $V_0$ to the shared \ml and forms a conducting path if the weight does \emph{not} match the respective multiplication result, realizing the \xnor operation.
Through the shared \ml, Kirchhoff's law accumulates the individual results.
The resulting current is proportional to the MAC value and charges the capacitor.
To reduce \spice simulation time, we use ideal Verilog-A implementations of the comparator and the FF (\SI{2}{\giga\hertz}).



\subsubsection{Framework Connecting PyTorch and SPICE}

Our framework loads the the information about the clippings (in CapMin in~\cref{eq:clipping}) and the error models (in CapMin-V in~\cref{eq:probm}) and applies them during the MAC computations of the BNNs in PyTorch.
In general, when the computing array and neuron circuit in \cref{fig:neuroncircuit} is used for computations, the MAC computations are separated into sub-MAC computations (see \cref{fig:errorappl}, in BNNs each matrix entry is binarized).
$\text{MAC}^1_1$ is the result of a vector product between the first row of the weight matrix $\mathbf{W}$ and the first column of the input matrix $\mathbf{X}$.
When it is computed with one computing array, the result of $\text{MAC}^1_1$ is summed up by sub-MAC results, i.e. $\text{MAC}^1_{1,1}$, $\text{MAC}^1_{1,2}$, etc.
To compute with $\beta$ weights and inputs, one array of size $a$ needs to be invoked $a_{last} = \lceil \frac{\beta}{a} \rceil$ times, and the results need to be summed, so that the entire value of $\text{MAC}^1_1$ can be computed.
In our framework, the error is applied at the level of these sub-MAC results.
In standard deep learning libraries (such as PyTorch), the sub-MAC results are not accessible.
To still enable to application of error models on the sub-MAC results, it is necessary to replace the closed source MAC engine with an own custom MAC engine.
We implemented our own custom MAC engine based on GPU CUDA kernel extensions for PyTorch in our framework.
Our custom CUDA-based MAC engine is called instead of the standard MAC engine of PyTorch.
With the full control over our custom MAC-engine, we equip it with functionality that enables clipping and error model application with any array size.
With that, arbitrary clippings and error models can be applied on the sub-MAC results.
Our framework is available as open-source in \url{https://github.com/myay/SPICE-Torch}.

\subsection{Minimizing Capacitor Size with Our Method CapMin}
\label{subsec:capminexp}

We extract \fmac by forward passes with the BNNs using the training datasets.
In \cref{fig:histograms} are the histograms of the absolute frequency of occurring MAC values.
Since all histograms are similar,
we normalize and add all the absolute frequencies across datasets and use the resulting \fmac in CapMin (\cref{subsec:m1}).

We apply CapMin using \fmac, $k$, and $a=32$ to obtain the set \smacmin.
In \cref{fig:accuracy_results}, the accuracy (test set) for different $k$ is shown in the plots (circle marks), starting with $k=32$ (max. nr. of levels for $a=32$) down to $k=5$.
For Fashion and Kuzujishi, the accuracy is sustained until $k=8$ and then drops sharply for smaller $k$.
For SVHN, CIFAR10, and Imagenette the accuracy is sustained until $k=8$ (SVHN) as well and $k=14$ (CIFAR10 and Imagenette) respectively.

In \cref{fig:cap_lat}, we show the reduction in capacitor size by CapMin.
The baseline has one spike time for each MAC level.
CapMin reduces the capacitor size by $14\times$, from \SI{135.2}{\pico\farad} to \SI{9.6}{\pico\farad}.
We show this in the bar plot with $k=14$, which can be used in a neuron circuit to achieve high accuracy in all five datasets.
This also leads to lower latency by $14\times$ as shown in \cref{fig:cap_lat}, where the guaranteed response time (GRT)~\cite{wei/etal/2021} is used to measure latency.
The energy reduction per MAC value computation is proportional to the capacitor size reduction, since the energy used in the capacitor is $\frac{1}{2}C\vth^2$, where $\vth=0.225$ \SI{}{\volt}.

\subsection{Increased Variation Tolerance with Our Method CapMin-V}
\label{subsec:capminVexp}
\vspace{-0.1cm}
When considering process variation
the current
has variations.
This varies the charging speed of the capacitor.
Consequently, the spike times may change, potentially leading to wrong MAC values.
To extract the error model (matrix $P_{\text{map}}$ in \cref{subsec:m2}), we use a Monte-Carlo approach (\num{1000} samples per spike time).
For each MAC level, a bucket is formed by placing decision boundaries midway between the spike times on either side.
The Monte-Carlo samples of a given MAC level are sorted into these buckets, counted, and the result normalized.
This yields the probability matrix $P_{\text{map}}$ to map from a MAC value to all possible MAC values.
Repeating this for all MAC levels constructs the entire mapping.
The errors from current variation are injected during the inference of BNNs and the average test accuracy of three runs is reported in the plots (star marks) in \cref{fig:accuracy_results}.
For all datasets the accuracy under variations is lower than without.
The accuracy drops are expected, due to the probabilities for wrong mappings.
Furthermore, the accuracy increases with smaller $k$.
This is due to the procedure of CapMin.
With smaller $k$, CapMin shifts the important spike times to more reliable, slower spike times.
In the case with e.g. $k=32$, the most reliable (longest) spike time is mapped to the highest MAC value.
In the case with $k=16$, the large and small MAC values are removed.
By this way, reliable spike times are mapped to important MAC values.
The sweet spots for $k$ and high accuracy under current variations are achieved for $15 \leq k \leq 12$ for Fashion, Kuzujishi, and SVHN, and for $15 \leq k \leq 14$ for CIFAR10.
We conclude that CapMin alone leads to variation tolerant operation to some extent.

However, under variations the accuracy drops for smaller $k$ compared to no variations.
We apply CapMin-V (\cref{alg:svmacmin}) to achieve higher tolerance to current variations.
For this, we use the capacitor size at $k=16$ (\SI{12.27}{\pico\farad}) with the corresponding \sfiremin as a starting point and evaluate for different $\phi$, so $\phi$ starts at $k=15$ and ends at $k=5$.
In \cref{fig:accuracy_results}, applying CapMin-V (triangle plots) sustains higher accuracy for more points compared to only CapMin (star plots).
In \cref{fig:cap_lat},
the capacitance in CapMin-V is merely $28\%$ and the latency $27\%$ larger compared to CapMin.
The capacitance (therefore energy) and latency are still $11\times$ smaller than the baseline.



\section{Conclusion}


We
propose CapMin, a HW/SW Codesign method for capacitor size minimization in analog computing IF-SNNs.
CapMin reduces the number of spike times needed in the HW based on MAC level occurrences in the SW.
Furthermore, we propose CapMin-V, which increases the tolerance to current variation.
CapMin achieves a 14$\times$ reduction in capacitor size over the state of the art, while CapMin-V achieves variation tolerance at small cost.
Our methods reduce area usage, energy, and latency, while increasing variation tolerance.

Our methods provide a cornerstone for exploring other NN models.
Despite the limitation of only using BNNs, the concepts can be extended to any NN models with higher precision.
However, analog computing with IF-SNNs for higher precision is a challenge due to the larger number of required analog states.
We plan to explore such extensions in the future work.




\bibliography{references}
\bibliographystyle{ieeetr}

\end{document}